\documentclass[multphys,vecphys]{svmult}

\usepackage{makeidx}         
\usepackage{graphicx}        
\usepackage{multicol}        
\usepackage[bottom]{footmisc}
\usepackage{epsfig}

\makeindex             

\begin{document}

\title*{A search for ultra-compact dwarf galaxies in the NGC 1023 group of galaxies}
\titlerunning{UCDs in the NGC 1023 group}
\author{Steffen Mieske\inst{1}\and
Michael J. West\inst{2}\and Claudia Mendes de Oliveira\inst{3}}
\institute{ESO, Karl-Schwarzschild-Str. 2, 85748 Garching b. M\"unchen,
Germany
\texttt{smieske@eso.org}
\and Department of Physics and Astronomy, University of Hawaii, Hilo, HI 96720
\texttt{westm@hawaii.edu}\and Instituto de Astronomia, Geof\'isica, e Ci\^encias Atmosf\'ericas, Departamento de Astronomia, Universidade de S\~ao Paulo, Rua do Mat\'ao 1226, Cidade Universit\'aria, 05508-090 S\~ao Paulo, SP, Brazil
\texttt{oliveira@astro.iag.usp.br}}
\maketitle

\begin{abstract}
We present a photometric search for UCD candidates in the nearby galaxy
group NGC 1023 (d=11 Mpc) -- the poorest environment searched for UCDs yet --, based on wide field imaging with CFHT. 
After photometric and morphological selection, we obtain a sample of 21 UCD candidates
with $-12<M_V<-11$ mag, if located at NGC 1023's distance. From spectroscopy taken at Calar Alto observatory, we identify the UCD candidate in closest
projection to NGC 1023 as an emission line background galaxy. Our photometric data show that in the NGC 1023 group, the mass spectrum of analogs to Fornax/Virgo UCD is restricted to 1/4 of
the maximum Fornax/Virgo UCD mass. More spectroscopy is needed to further
constrain the mass range of UCDs in this galaxy group.
\end{abstract}

\section{Introduction}
In their spectroscopic studies of the Fornax cluster of galaxies, 
\cite{mie_Hilker99}$-$\cite{mie_Drinkw03} discovered six 
isolated compact stellar systems close to the cluster centre with luminosities $-13.4<M_V<-12$ mag and half-light radii $r_h$ between 15 and 30 pc. They were dubbed ``ultra-compact dwarf galaxies`` (UCDs) 
(\cite{mie_Philli01}). In \cite{mie_Mieske06}, metallicities and structural parameters for compact objects in Fornax were derived, indicating that the realm of UCDs in Fornax indeed extends down to about $M_V\simeq -11$ mag. 
UCDs have also been found in the Virgo and Abell 1689 clusters (\cite{mie_Hasega05}$-$\cite{mie_Mieske05a}).\\
Two main formation scenarios for these puzzling objects have been brought forward. 1. UCDs are tidally stripped ``naked'' nuclei of dE,Ns (\cite{mie_Bekki03}). 2. UCDs are merged clusters of massive star clusters created in gas-rich galaxy-galaxy mergers (\cite{mie_Fellha02}).\\
In order to improve our understanding of UCDs, it is necessary to define their properties in a range of host environments.\\
In this contribution, we report on a search for UCDs in the nearby galaxy group NGC 1023. This is the poorest environment that has been searched for UCDs, yet. The galaxy NGC 1023 is of type SB0, has $M_R=-21$ mag (\cite{mie_Trenth02}), and has a distance modulus of 30.29 mag (\cite{mie_Tonry01}). The NGC 1023 group consists of a few dozen mainly late type galaxies (\cite{mie_Trenth02},\cite{mie_Tully80}). It has a  narrow velocity distribution with $\sigma \simeq$ 55 km/s and a virial mass of about 2$\times$10$^{12} M_{*}$. 
\section{A search for UCDs in NGC 1023}
\subsection{Data and selection criteria}
\label{mie_photsel}
\begin{figure*}
   \begin{center}
   \epsfig{figure=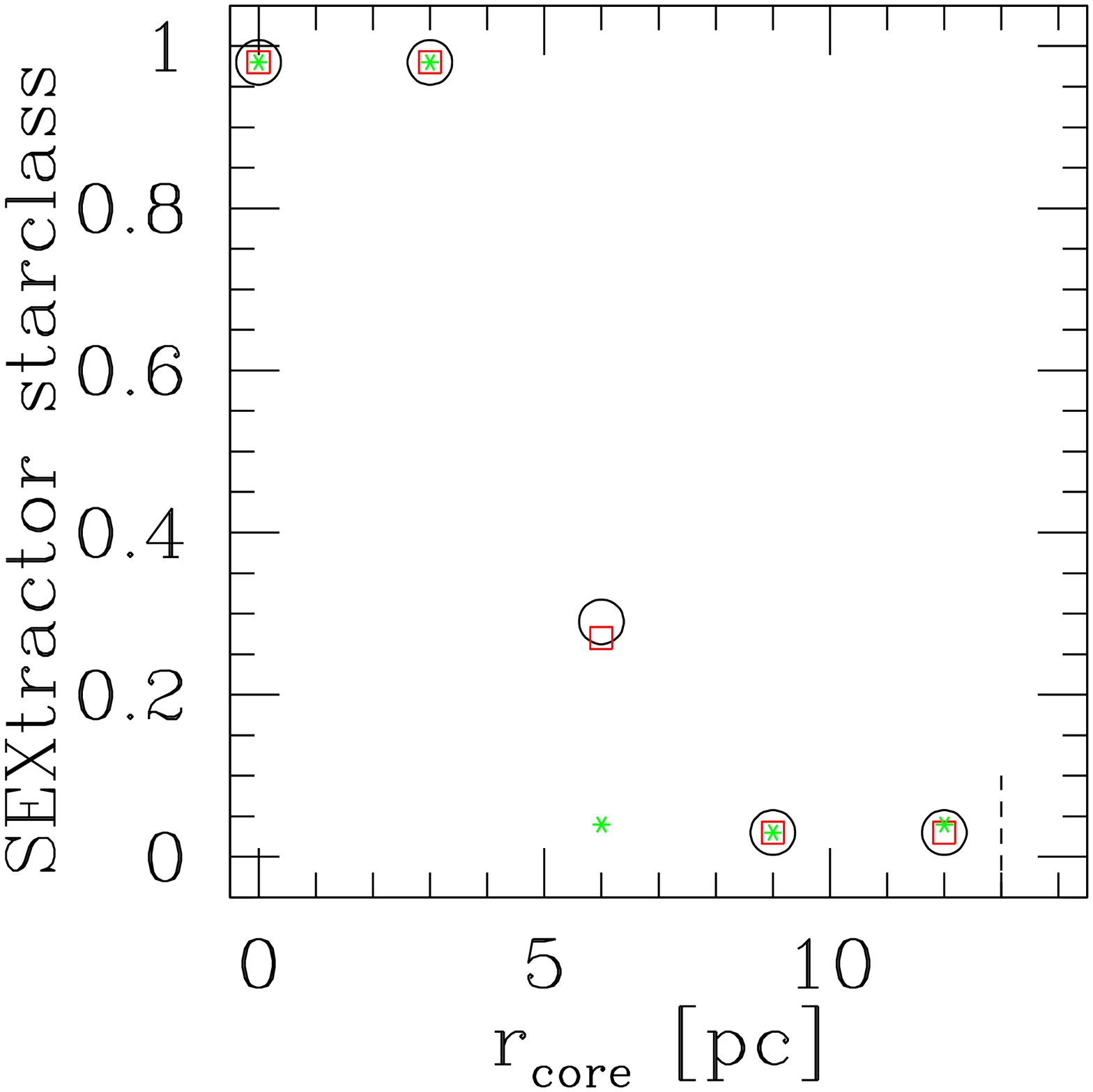,width=3.7cm}\hspace{0.15cm}
\epsfig{figure=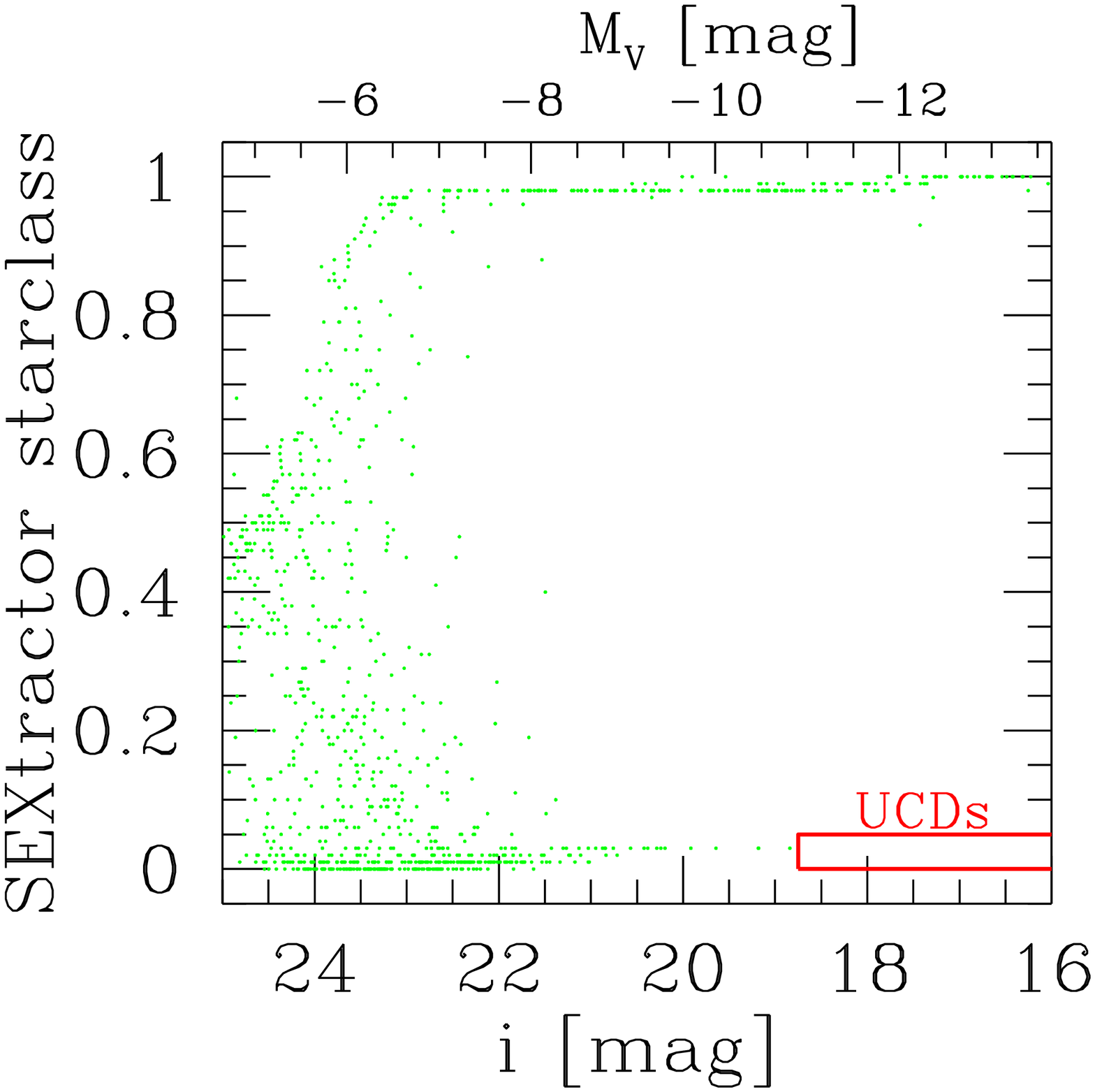,width=3.7cm}\hspace{0.15cm}
\epsfig{figure=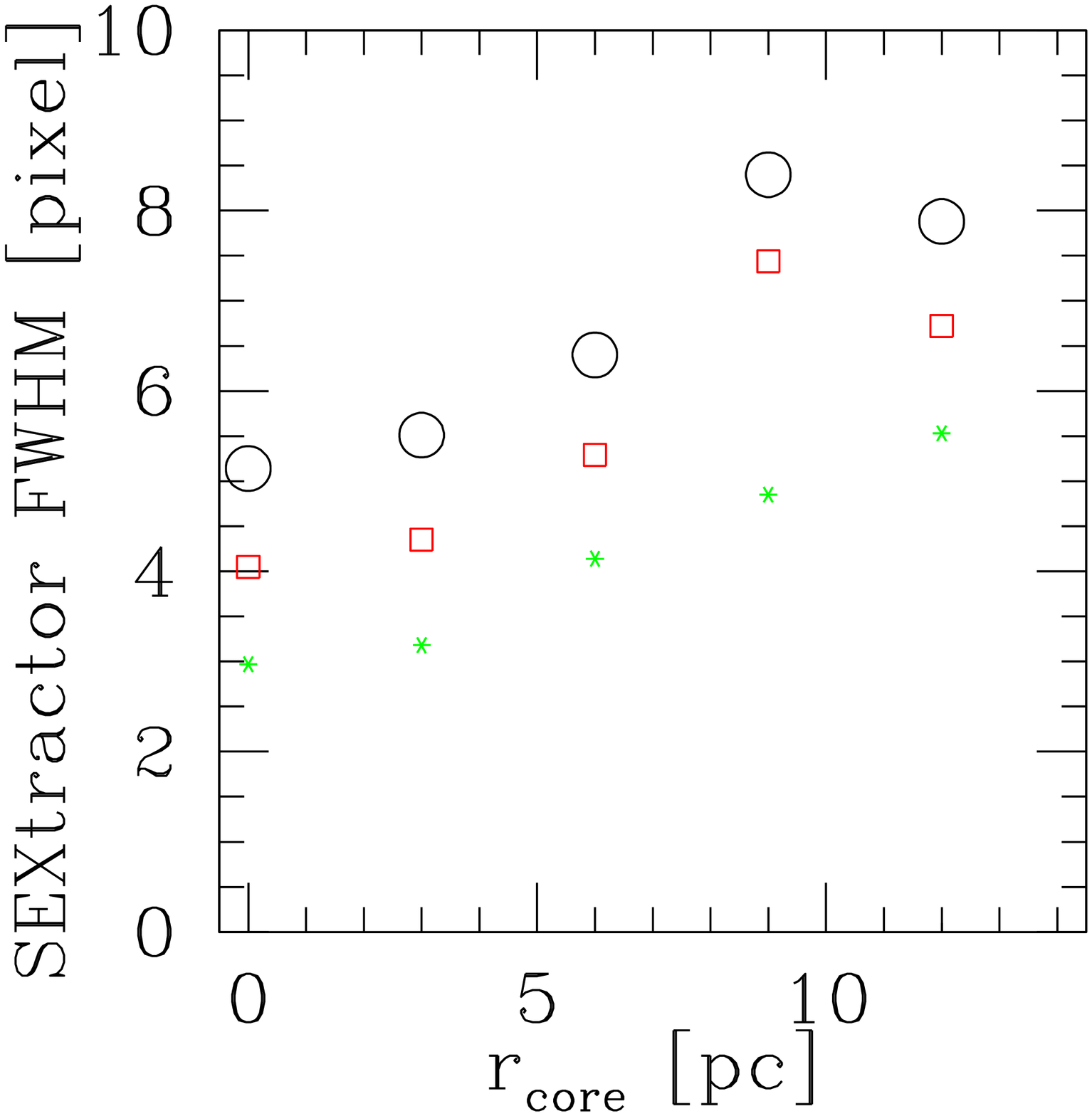,width=3.7cm}
      \caption{{\bf Left:} SExtractor star-classsifier plotted vs. core radius $r_{core}$ for simulated UCD-type sources in our CFHT images, projected to NGC 1023's distance. The assumed intrinsic UCD profile is a King profile with concentration parameter c=1.2 (\cite{mie_Hasega05}). The profile is convolved with three different PSFs: large black circles denote FWHM=0.9$''$; medium sized red squares denote FWHM=0.7$''$; small sized green asterisks denote FWHM=0.5$''$. The dashed tick at r=13 pc denotes the mean UCD King core radii from \cite{mie_Hasega05}. {\bf Middle:} SExtractor star-classifier plotted vs. apparent $i$ magnitude for 1 out of the 144 imaged CCD chips. Objects plotted are pre-selected according to the UCD colour, size and ellipticity criteria outlined in Sect.~\ref{mie_photsel}. The rectangle at the lower right corner of the panel indicates the magnitude+starclass selection criteria for UCD candidates. {\bf Right:} Plot is analogous to the one on the left - here the y-axis displays the SExtractor FWHM in pixels instead of the SEXtractor starclass. }
         \label{mie_sexclass}
\end{center}
   \end{figure*}

\begin{figure*}[ht!]
   \begin{center}
   \epsfig{figure=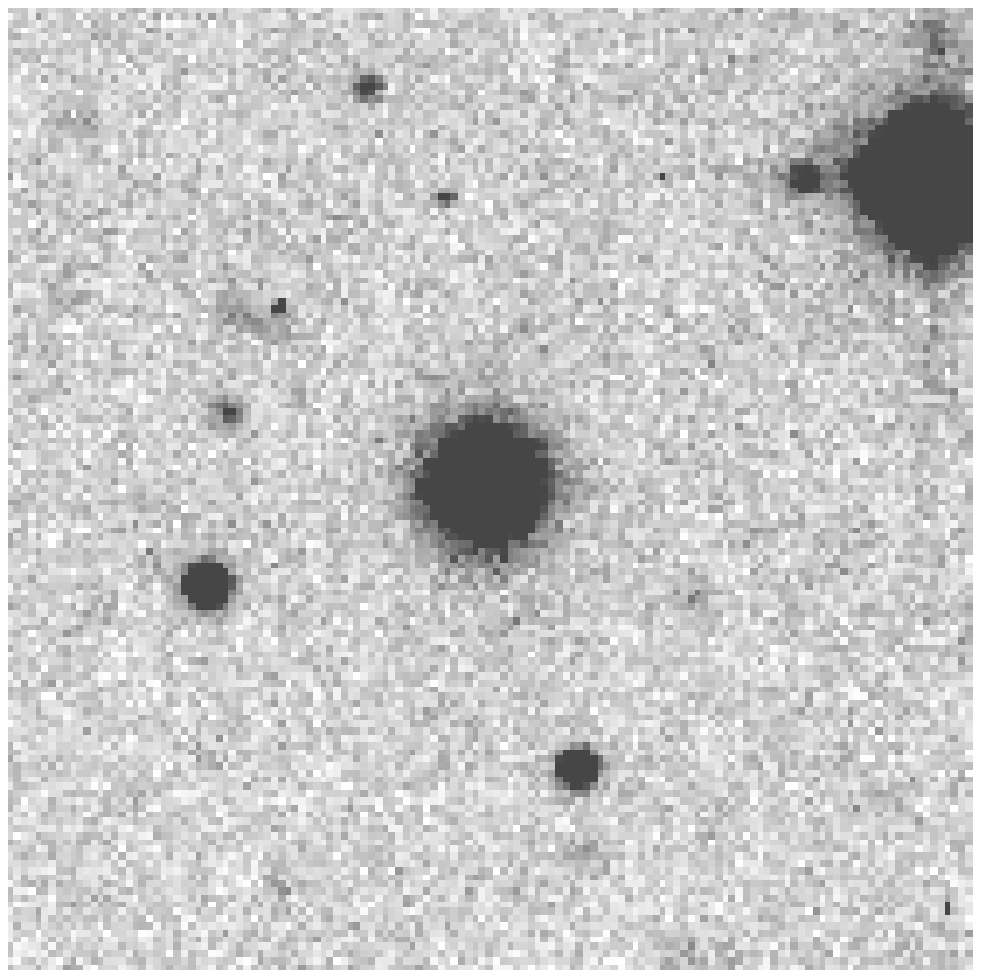,width=3.6cm}\hspace{0.15cm}
\epsfig{figure=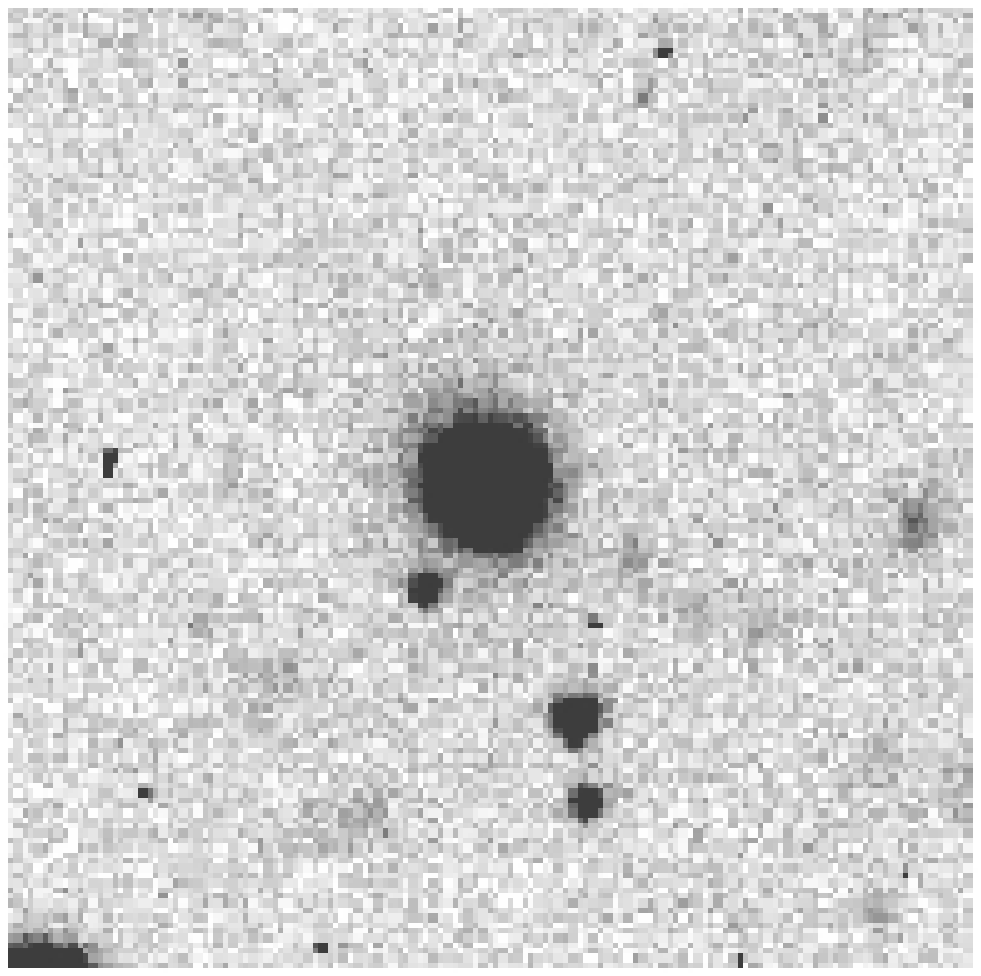,width=3.6cm}\hspace{0.15cm}
\epsfig{figure=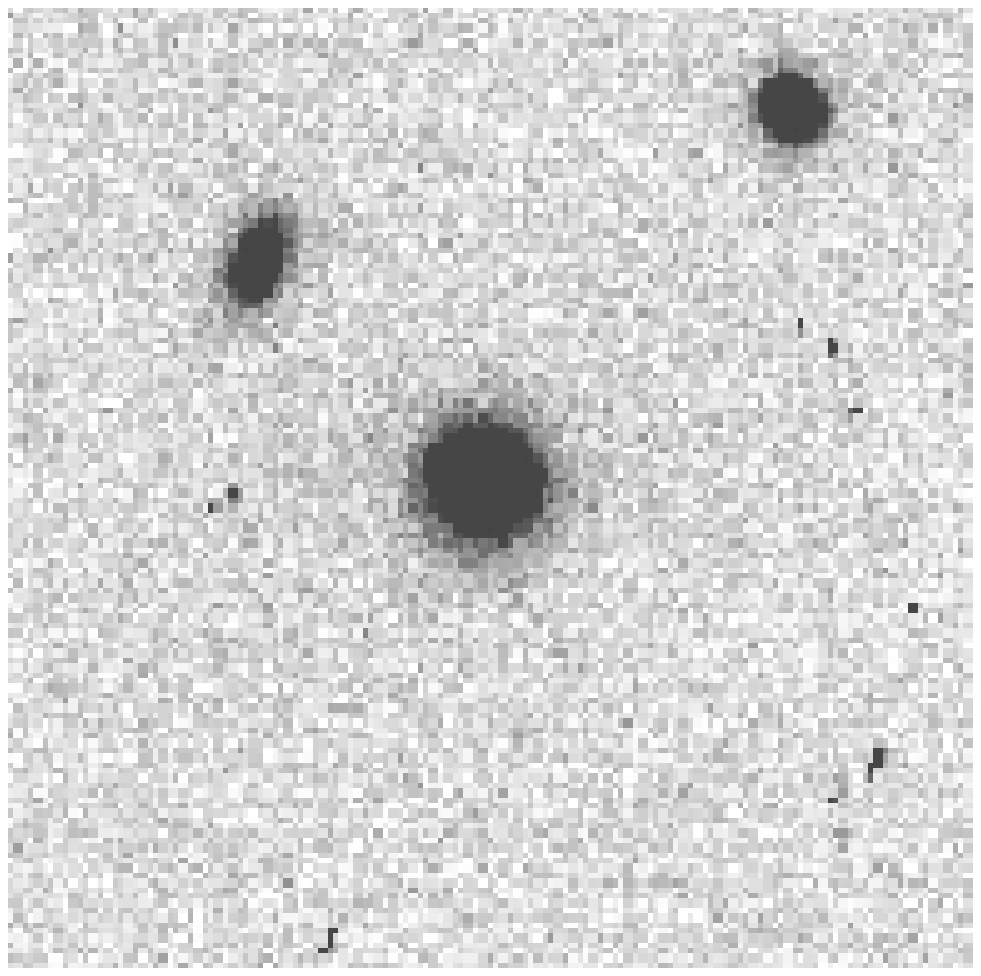,width=3.6cm}
\epsfig{figure=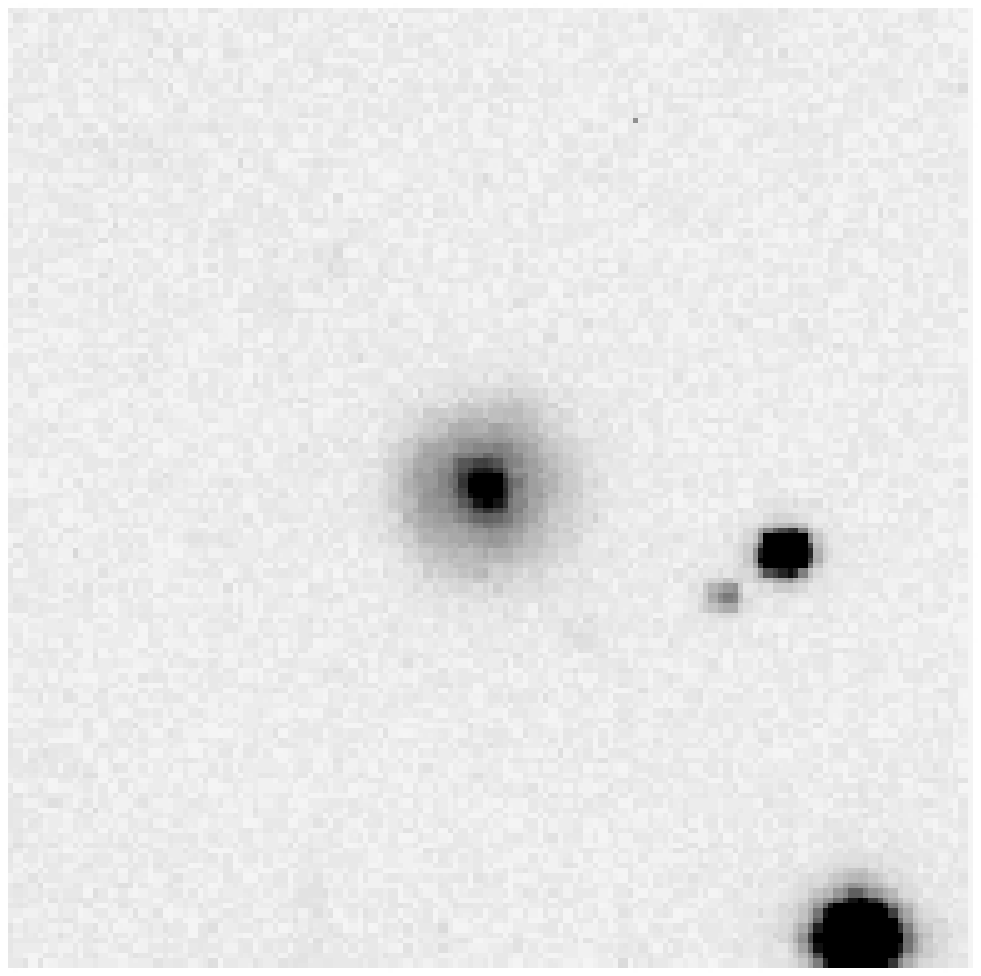,width=3.6cm}\hspace{0.15cm}
\epsfig{figure=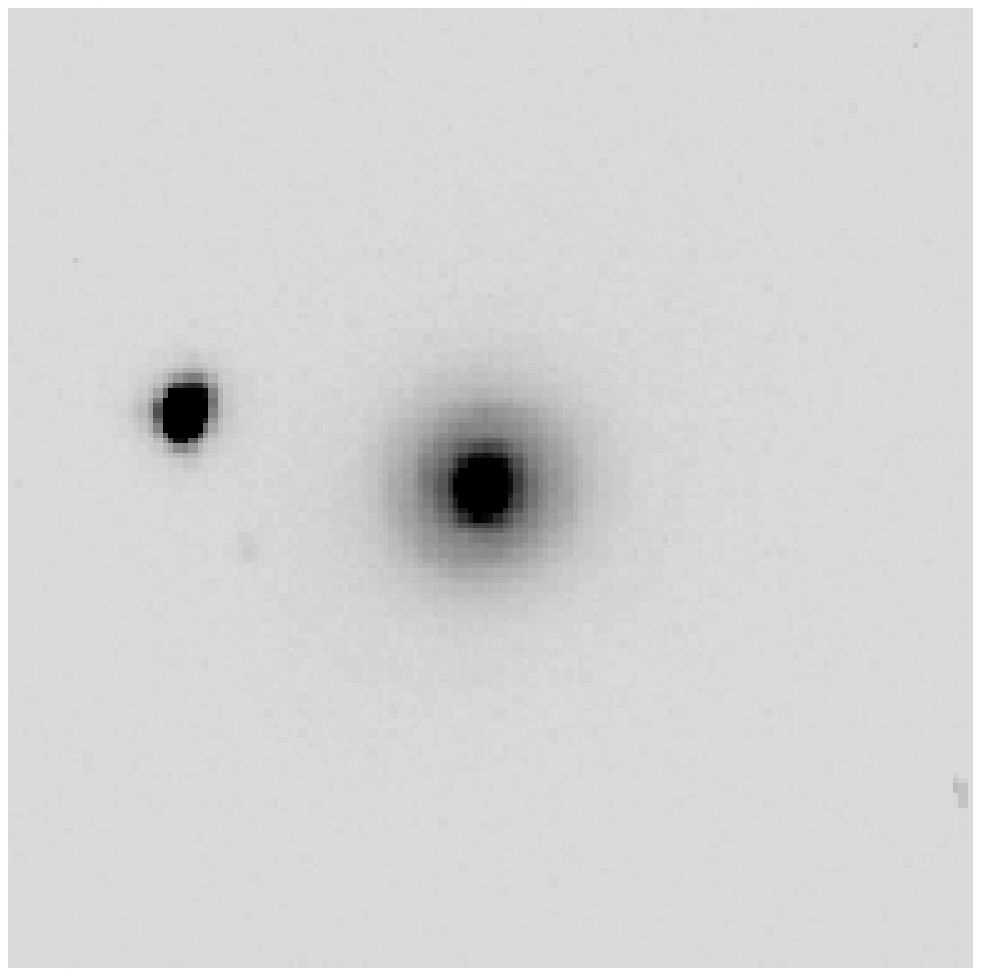,width=3.6cm}\hspace{0.15cm}
\epsfig{figure=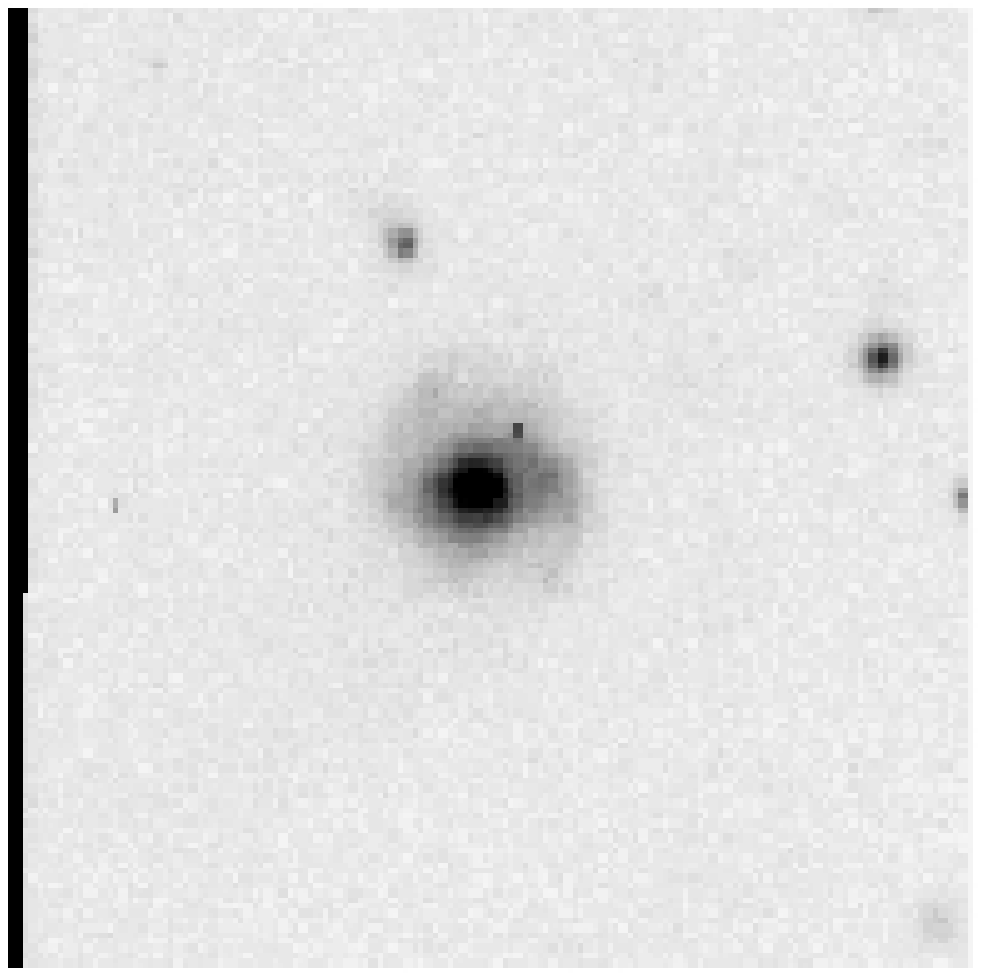,width=3.6cm}
      \caption{Thumbnail images of 26$\times$26$''$ illustrating the morphological selection of UCD candidates. {\bf Top row:} 3 examples for photometrically pre-selected SExtractor detections that morphlogically classify as UCD candidates because of their featureless compactness. {\bf Bottom row:} 3 examples for photometrically pre-selected SExtractor detections that morphologically classify as bulges or nuclear regions of background spirals.}
         \label{mie_images}
\end{center}
   \end{figure*}

\begin{figure*}[h!]
\begin{center}
   \epsfig{figure=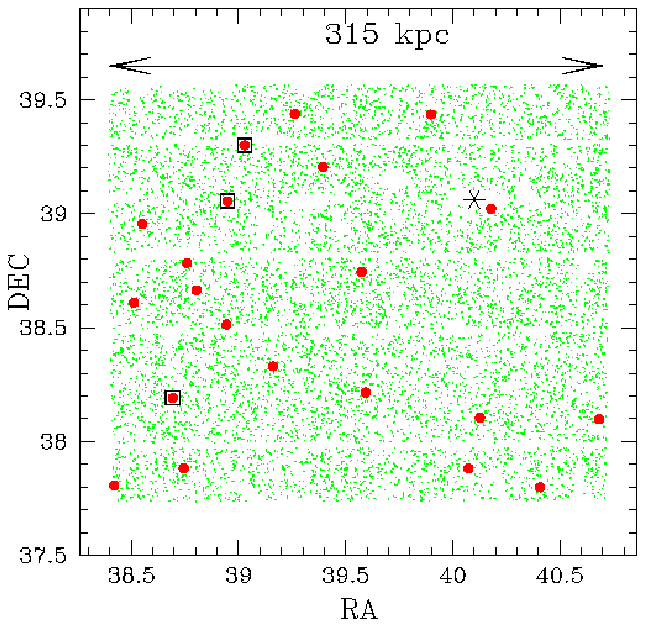,width=5.72cm}\hspace{0.15cm}
\epsfig{figure=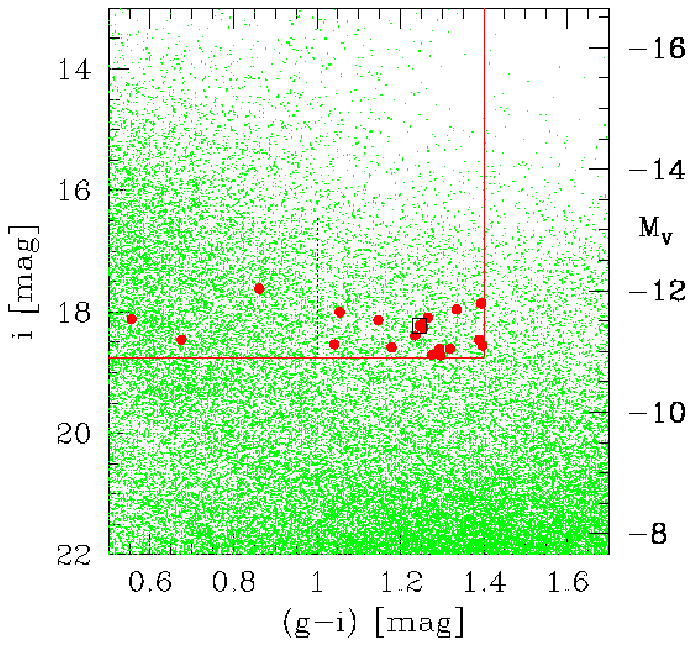,width=5.72cm}
      \caption{{\bf Left:} Map of the investigated area. Red (large) dots indicate positions of UCD candidates. Green (small) dots are pre-selected UCD candidates based only on colour and magnitude. The position of NGC 1023 is indicated by the asterisk. The three UCD candidates marked by squares are those blue-wards of the Fornax/Virgo UCD colour range as indicated in the right panel. {\bf Right:} Green (small) dots show a colour-magnitude diagram of all detected sources in the surveyed area.
 The solid red lines denote the colour-magnitude selection criteria for UCD candidates.
The dashed lines denote the colour-magnitude range of Fornax/Virgo UCDs. The (large) red circles show the UCD candidates. The one source marked by a square is UCD candidate f1\_15, which is in closest projected distance to NGC 1023, see left panel. }
         \label{mie_mapcmd}
\end{center}
   \end{figure*}

The imaging data for this investigation were obtained at the 3.6m Canada-France-Hawaii-Telescope (CFHT) at Mauna Kea in service mode between September and November 2004, using the wide-field imager MegaPrime / MegaCam. This camera images a field-of-view of 1$\times$1 degree onto 36 2048 x 4612 pixel CCDs, which have a pixel scale of 0.187$"$. Four fields around NGC 1023 were imaged (see Fig.~\ref{mie_mapcmd}) in $g$ and $i$. The seeing-FWHM was between 0.6 and 0.9$''$.
The advantage of the NGC 1023 imaging data is that UCD analogs to those in Fornax and Virgo are resolved, see the Monte Carlo simulations in Fig.~\ref{mie_sexclass}. This makes the photometric UCD candidate search very effective.\\
We applied the following criteria to select UCD candidates:\\
{\bf 1. i$<$18.75 mag.} At the distance of NGC 1023, this corresponds to $M_V=-11$ mag (\cite{mie_Fukugi96}, \cite{mie_Schleg98}), which is the approximate limit between UCDs and GCs (\cite{mie_Mieske06}). \\
{\bf 2. (g-i)$<$1.40 mag.} This corresponds to the red limit $(V-I)_0\simeq1.3$ of Fornax UCDs (\cite{mie_Mieske06}, \cite{mie_Fukugi96}, \cite{mie_Schleg98}, \cite{mie_Bruzua03}). \\
{\bf 3. Ellipticity $<$ 0.2.} None of the UCDs resolved with HST imaging shows ellipticities larger than 0.2 (\cite{mie_Drinkw03},\cite{mie_Hasega05}).\\
{\bf 4. SExtractor FWHM $<15$ pixel.} This corresponds to $\simeq$ 150 pc at NGC 1023's distance and is chosen to reject significantly extended background sources or cluster members. UCDs are not expected to surpass this size limit, see the Monte Carlo simulations in the right panel of Fig.~\ref{mie_sexclass}. \\
{\bf 5. SExtractor Star Class $<$ 0.05.} This limit is applied because analogs to Fornax/Virgo UCDs will be resolved, see Fig.~\ref{mie_sexclass}. It biases us against detecting UCDs smaller than core radii of about 8pc, which is the very lower limit of UCD sizes in Fornax and Virgo.\\
{\bf 6. Morphological selection.} 
The final step was to morphologically reject bulges / nuclear regions of background galaxies, which constitute the major part of the pre-selected sample (Fig.~\ref{mie_images}). About 90\% of the photometrically pre-selected objects were classified by visual inspection as background galaxies, leaving 21 bona-fide UCD candidates.

\subsection{Results}
\begin{figure}[ht!]
\begin{center}
   \epsfig{figure=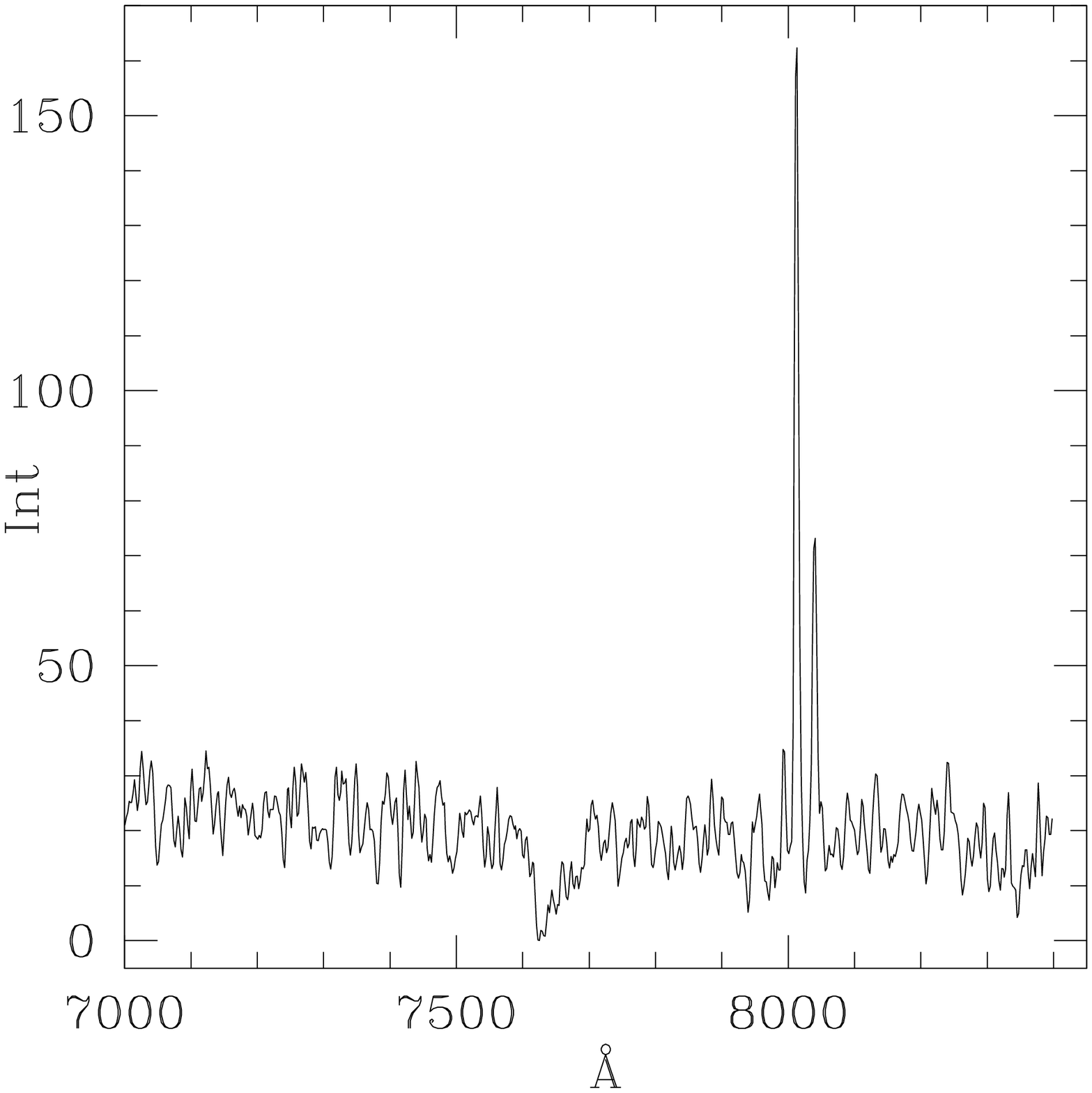,width=6.5cm, height=5.8cm}\hspace{0.15cm}
      \caption{Excerpt of the spectrum of UCD candidate f1\_15, which is the one in closest projection to NGC 1023. We identify the typical Si telluric absorption band at $\simeq$ 7620 $\AA$. The group of three emission lines is identified (in order of decreasing equivalent width) as H$\alpha$ @ rest-frame wavelength 6562.8 $\AA$, NII @ 6585 $\AA$ and NII @ 6549 $\AA$. The corresponding redshift of the source is then z=0.221.}
         \label{mie_f1_15}
\end{center}
   \end{figure}

The final sample of 21 bona-fide UCD candidates has $-12<M_V<-11$ mag and is shown in a map and colour-magnitude diagram of the surveyed area in Fig.~\ref{mie_mapcmd}. 
The most striking feature of this final UCD candidate sample is that there are no UCD candidates with $M_V<-12$ mag. That is, the mass spectrum of possible UCDs in the NGC 1023 group is restricted to about four times lower masses than that of Fornax / Virgo UCDs. This confirms the intuitive expectation that both the number and mass of UCDs scale with host environment properties such as galaxy density and depth of potential well. For example, there are no UCDs known in the Local Group. The spatial distribution of the UCD candidates is clearly not centered on NGC 1023, consistent with a substantial background contamination.\\
We have obtained a spectrum for the UCD candidate in closest projection to NGC 1023, called f1\_15. We were granted 1 hour of observation in DDT mode at Calar Alto observatory using CAFOS (Calar Alto Focal Reducer and Faint Object Spectrograph) mounted at the 2.2m telescope with the G 100 grism (instrumental resolution 7 $\AA$). From the spectrum of f1\_15 (Fig.~\ref{mie_f1_15}) we identify it as an emission line galaxy at z=0.211.
\section{Conclusions}
Using only photometric data we have been able to show that in the NGC 1023 group the mass spectrum of Fornax/Virgo UCD analogs does not extend beyond about 1/4 of
the maximum UCD mass in Fornax/Virgo. 
More spectroscopy will be needed to further constrain the mass range of possibly existing UCDs in this group.
\label{mie_conclusions}

\printindex
\end{document}